# Supersonic Combustion Diagnostics with Dual Comb Spectroscopy

David Yun[a,*], Nathan A. Malarich[a], Ryan K. Cole[a], Scott C. Egbert[a],
Jacob J. France[b], Jiwen Liu[c],
Kristin M. Rice[d], Mark A. Hagenmaier[d], Jeffrey M. Donbar[d],
Nazanin Hoghooghi[a], Sean C. Coburn[a], Gregory B. Rieker[a,**]

[a]*Precision Laser Diagnostics Laboratory, University of Colorado Boulder, Boulder, CO 80309, USA*
[b]*Innovative Scientific Solutions Incorporated, Dayton, OH 45459, USA*
[c]*Taitech Incorporated, Beavercreek, OH 45430, USA*
[d]*U.S. Air Force Research Laboratory, Wright-Patterson AFB, OH 45433, USA*

___________________________________________________________________________________

**Abstract**

Supersonic engine development requires accurate and detailed measurements of fluidic and thermodynamic parameters to optimize engine designs and benchmark computational fluid dynamic (CFD) simulations. Here, we demonstrate that dual frequency comb spectroscopy (DCS) with mode-locked frequency combs can provide simultaneous absolute measurements of several flow parameters with low uncertainty across a range of conditions owing to the broadband and ultrastable optical frequency output of the lasers. We perform DCS measurements across a 6800-7200 $cm^{-1}$ bandwidth covering hundreds of $H_2O$ absorption features resolved with a spectral point spacing of 0.0067 $cm^{-1}$ and point spacing precision of $1.68 \times 10^{-10}$ $cm^{-1}$. We demonstrate 2D profiles of velocity, temperature, pressure, water mole fraction, and air mass flux in a ground-test dual-mode ramjet at Wright-Patterson Air Force Base. The narrow angles of the measurement beams offer sufficient spatial resolution to resolve properties across an oblique shock train in the isolator and the thermal throat of the combustor. We determine that the total measurement uncertainties for the various parameters range from 1% for temperature to 9% for water vapor mole fraction, with the absorption database/model that is used to interpret the data typically contributing the most uncertainty (leaving the door open for even lower uncertainty in the future). CFD at the various measurement locations show good agreement, largely falling within the DCS measurement uncertainty for most profiles and parameters.

___________________________________________________________________________________

*Corresponding author. david.yun@colorado.edu

**email address: greg.rieker@colorado.edu

## 1. Introduction

Supersonic systems are an important topic of research and development for various applications such as planetary reentry and national defense. Experimental and simulation-based evaluation of supersonic vehicles requires accurate and detailed diagnostics of various fluidic and thermodynamic parameters across vastly different conditions within and around the vehicle. For example, in supersonic air-breathing ramjet and scramjet engines, both inlet mass flux and complex flow conditions in the flame holding region can impact flight performance and even prevent combustion reactions [1–5]. In this study, we use dual-frequency comb spectroscopy (DCS) with mode-locked frequency combs to make accurate, spatially resolved measurements of multiple flow parameters in both the relatively low-temperature isolator and the high-temperature combustor of a ground-test dual-mode ramjet engine.

DCS [6,7] is an emerging form of laser absorption spectroscopy (LAS). LAS involves passing a beam of light through a gas sample and measuring the absorption that occurs at wavelengths resonant with the rotational-vibrational transitions of the molecules in the flow. Flow velocity, pressure, temperature, and species mole fraction (and consequently other flow parameters such as mass flux) can be simultaneously and absolutely determined by matching the shape, size, and position of features in the measured absorption spectra to physical absorption models [8–15].

DCS with mode-locked frequency combs provides two attractive qualities for low-uncertainty LAS in supersonic environments: broad optical bandwidth and an extremely precise and stable optical frequency axis [16]. Mode-locked frequency combs emit an optical pulse train composed of broadband light spanning an optical bandwidth that is tens to hundreds of times wider than narrowband forms of laser absorption spectroscopy (e.g tunable-diode LAS), and made up of frequency elements (comb teeth) that are perfectly spaced by the pulse repetition rate of the laser. The broad bandwidth allows the absorption model fitting algorithm to utilize hundreds of individual absorption features to determine fluidic and thermodynamic properties with large dynamic range and robustness against optical interference effects such as etalons and baseline laser intensity fluctuations. The optical frequency axis stability improves measurement accuracy and precision of the relative positions and widths of absorption features, which are essential parameters for pressure and velocity retrievals, as demonstrated for velocity in [17].

In a prior publication, we demonstrated spatially resolved mass flux measurements using dual-comb spectroscopy and proved via comparison with simple, known conditions and careful analysis that the approach can reach low levels of uncertainty [16]. Here, we take advantage of the precision and low uncertainty to narrow the spatial extent of the probe laser beams (which are angled to enable velocimetry) and resolve a 2D profile of velocity, temperature, pressure, $H_2O$ mole fraction ($\chi_{H_2O}$), and mass flux across an oblique shock train in a ground-test dual-mode ramjet isolator. We also demonstrate the first DCS in a high-speed combustor, measuring the 2D profile of velocity across the thermal throat under harsh conditions. We compare computational fluid dynamics (CFD) calculations to the measured profiles, finding that a majority of the CFD values are in reasonable agreement, including those in the combustor. We perform a thorough uncertainty analysis for each of these measurements, considering all major sources of uncertainty. We find the measurement uncertainties range from 1% to 9%, with the potential for improvement. The results demonstrate dual-comb spectroscopy as a versatile supersonic combustion diagnostic.

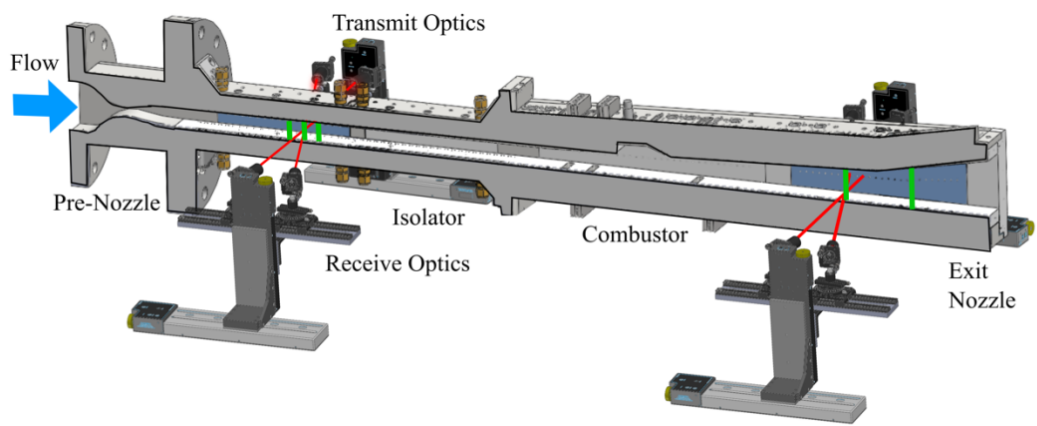

Fig. 1. RC-18 ground-test, direct-connect, dual-mode ramjet test engine with dual-comb optics. The pre-nozzle accelerates vitiated air to simulate in-flight inlet conditions. The isolator compresses air via an oblique shock train that is created by a distortion generator (not shown). The combustor injects and ignites fuel and the exit nozzle accelerates the heated air. DCS light is transmitted across the engine in an angled dual beam configuration. Motor stages move optics to take vertical scans (indicated by green lines) at different axial locations to create 2D profiles in the isolator and near the combustor thermal throat.

## 2. Methods

We perform the measurements at Wright-Patterson Air Force Base Research Cell 18 [18] in a ground-test, direct-connect dual-mode ramjet (Fig. 1). To replicate the supersonic flow entering the inlet of an in-flight vehicle, air is vitiated and accelerated through a nozzle before proceeding to the test section. The air then flows through the isolator, where a distortion generator (not shown in Fig. 1) creates an oblique shock train that reflects repeatedly from top to bottom of the isolator flow. The post-shock-train air then enters the combustor at an elevated temperature and pressure and a lower velocity. Within the combustor, fuel is injected and combusted in the flame-holding region, shown in the top wall of the combustor in Fig. 1. Finally, the diverging exit nozzle accelerates the combustion products to produce thrust.

To interrogate the system, we use DCS light which has several qualities that are well-suited for measurements in dynamic systems [19–26] and has been previously demonstrated in several combustion environments [27–29]. The mode-locked, erbium-doped fiber frequency comb lasers used in this study [30] produce light that allows for measurements with an optical bandwidth of >400 cm$^{-1}$ which encompass thousands of $H_2O$ absorption features. This broadband light consists of tens of thousands of tightly and equally spaced optical frequency elements (comb teeth) which constitute a spectrum that finely resolves each of the individual absorption features. The spectral point spacing of the mode-locked frequency combs in this study is 0.0067 cm$^{-1}$ or 200 MHz. The spacing is set by the pulse repetition rate of the combs, which is continuously controlled to 2.5 parts in 10$^8$, lending an overall optical frequency spacing precision of $1.68 \times 10^{-10}$ cm$^{-1}$ [16].

In DCS, two combs with slightly different pulse repetition rates (i.e. frequency spacings) are interfered to produce radio frequency beat notes that can be easily read by a fast photodetector for each comb tooth pair [6]. Full spectra are acquired at a rate equal to the difference of the pulse repetition rates of the two combs (626 Hz for this study). Spectra are then coherently averaged to reach the desired signal-to-noise ratio. We average for 30 seconds in the isolator and 45 seconds in the combustor to reach sub-2% precision for most measured quantities and sub-4% for all quantities, as described later.

To take absorption measurements in the test engine, light from the DCS system goes through an optical filter that selects the desired wavelength range of the frequency comb light. Different wavelength ranges are used for the isolator and combustor as shown in Fig. 2. The light then splits onto two paths of single-mode optical fiber using a 50/50 fiber coupler. The light on each path launches through the test engine quartz optical access windows via collimating transmit optics. Rotation stages angle the paths in an upstream-propagating and downstream-propagating direction (hereafter, just referred to as the upstream and downstream). The precision of DCS allows us to use more narrow angles, 12.5° in the isolator and 15° in the combustor relative to the normal of the access windows, compared with past LAS velocimetry studies. These angles give the crossed beams a 2.2 and 2.8 cm streamwise spatial extent in the isolator and combustor flow respectively. The transmit and receive setups are each placed on vertical and horizontal stages to control the measurement location and enable 2D mapping of the flow parameters. Light is received onto multimode fiber on the other side of the isolator, and the DCS

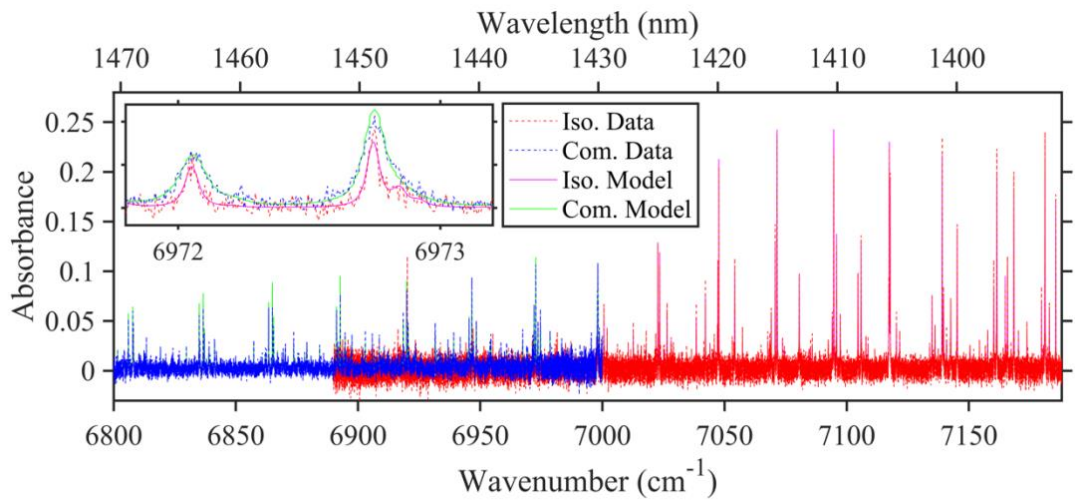

Fig. 2. Spectra taken in the isolator (red) and the combustor (blue). Isolator spectra are averaged for 30 s and combustor spectra in the combustor were averaged for 45 s. The inset shows a zoomed region with model fits.

signals are measured on fast photodetectors and recorded with two 250 MS/s digitizers.

## 3. Experiment 1: 2D Multi-parameter Profiles in the Isolator

The oblique shock train in the isolator provides a challenging environment for diagnostics. The shock train creates sharp gradients in multiple flow parameters along the vertical and horizontal directions, which wall-mount pressure and temperature measurements and intrusive diagnostics cannot resolve. We take DCS spectra at various heights and axial locations to spatially resolve these gradients. For the isolator measurements, we trim the broadband DCS light with optical filters to span 6880 – 7186 cm$^{-1}$ as shown in Fig. 2. This region spans a high-temperature water ($H_2O$) spectroscopic absorption database with pure water parameters from Schroeder et al. [31] and air-broadening parameters from HITRAN2012 [32] which are validated at the temperature and pressure conditions found in the isolator [33]. Velocity, pressure, temperature and $H_2O$ mole fraction are retrieved by fitting spectra from the upstream and downstream paths to absorption models built from the database [34].

The bulk velocity of the flow in the isolator produces a frequency shift in the absorption feature positions induced by the Doppler effect. The equal but opposite angle between the upstream and downstream laser beam and the flow result in equal but opposite Doppler shifts. We use the same crossed-beam configuration as employed in similar LAS velocimetry studies [8–12] to decrease velocity measurement uncertainty. Here, the angles are set to +12.5° and -12.5°, narrower than most LAS velocimetry to date, which allows the measurements to resolve centimeter-scale streamwise gradients in the rapidly changing flow.

Pressure correlates to the measured width of the absorption features, temperature to the integrated absorption area ratios between features, and $H_2O$ mole fraction to the total light absorbed across the spectra. Mass flux is calculated from the velocity, pressure, temperature, and $H_2O$ mole fraction values using a derivation of the ideal gas law [16].

We take DCS measurements at three axial positions downstream of the isolator entrance (220, 240, and 260 mm). At each axial position we scan the laser beams in 1-2 mm vertical increments spanning the isolator. The measurement locations are represented by the vertical green lines in Fig. 1 and create a 2D profile of mass flux along an oblique shock. DCS measurements that do not meet an SNR threshold are not included in this study. These were all near the top or bottom wall of the isolator where a turbulent boundary layer likely contributed to a drop in laser throughput due to beam steering.

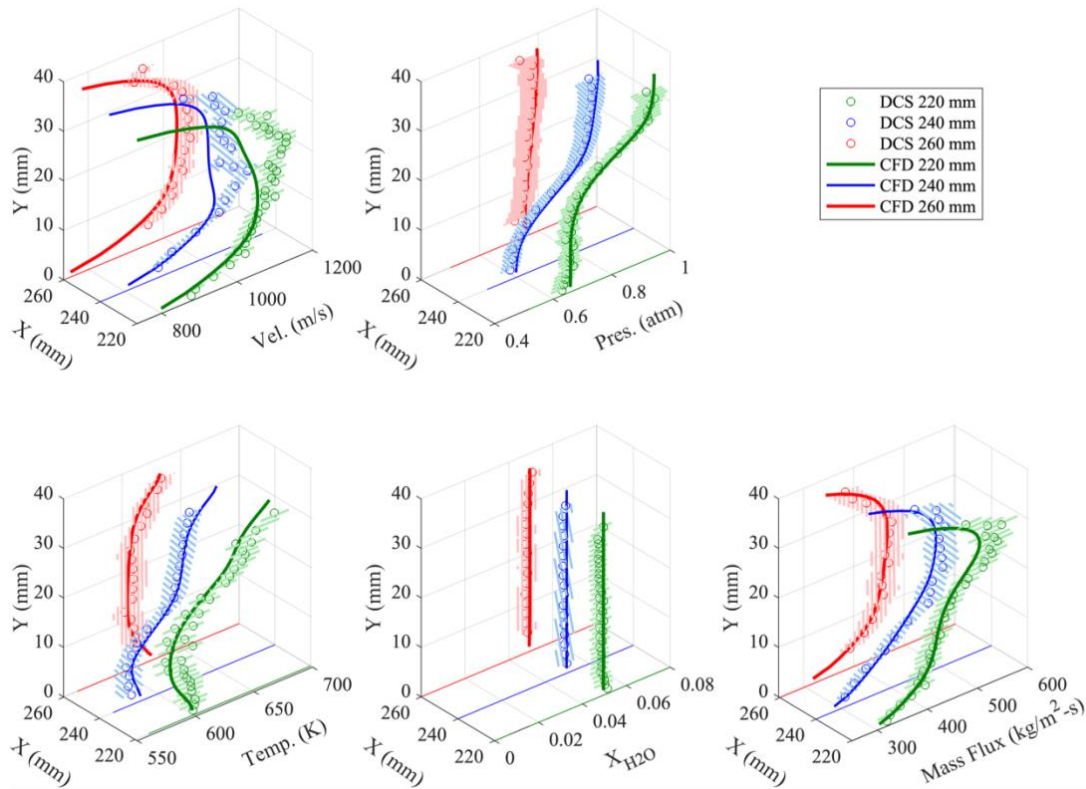

Fig. 3. Comparison of DCS-measurements (markers) and CFD (lines) of velocity, temperature, pressure, $\chi_{H_2O}$, and mass flux at various heights (Y) and axial locations (X) in the isolator. Zero height is defined as the bottom of the isolator and zero axial position is defined as the isolator entrance. Hatched regions indicate the DCS measurement uncertainty bounds.

The DCS results are shown in Fig. 3. They show many interesting characteristics of the flow. The velocity slows near the top and bottom walls, and there is a corresponding increase in the static temperature. The water vapor mole fraction is nearly constant across the entire vertical and axial extent of the measurements, as expected since water vapor is neither produced nor destroyed through the isolator. The vertical pressure gradients for the three axial locations show the changing oblique shock location. The measurements cross the shock near the top wall at the 220 mm location, near the middle of the duct at the 240 mm location, and near the bottom wall at the 260 mm location.

Uncertainties were calculated for each measured quantity by considering all major sources of uncertainty using the same method described in [16]. An example of the uncertainty analysis can be seen in Table 1 for the measurement at the 240 mm axial and 16 mm height location. Calculated uncertainties for this measurement are 3.0% for velocity, 5.6% for pressure, 1.9% for temperature, 8.5% for $\chi_{H2O}$, and 6.6% for mass flux. The uncertainty for other measurement points in the isolator are similar. The analysis shows that the largest sources of uncertainty are the precision (which stems from the measurement noise due to sensor noise or engine fluctuations at 30s averaging time), beam angle, and database uncertainty. These sources (and thus the overall measurement uncertainty) can be improved by longer averaging times, improvements to optical alignment, and updates to spectroscopic databases, respectively. A detail discussion of the uncertainty for multi-parameter DCS measurements can be found in [16].

Comparisons of the measurements with CFD model values are also shown in Fig. 3. The CFD is made in CFD++ (Metacomp Technologies, Inc.) using a full 3D Reynolds-Averaged Navier Stokes. Turbulence is modeled using a two equation cubic $k - \epsilon$ model. The turbulent Prandtl number is set to 0.9 and the Schmidt number is calibrated from experimental pressure tabs along the engine. At the walls, a two-layer wall function with blended equilibrium and non-equilibrium modes are employed to reduce grid requirements. To directly compare the CFD values to the laser-measured values (which are line-of-sight averaged), we simulate a laser measurement through the CFD environment and calculate a path-integrated simulated spectra which we fit with the same fitting routines used on the real laser measurements. This allows the CFD-derived values to be affected by the same small spectral fit biases that the DCS measurements may experience due to flow non-uniformities [35] thus allowing for a more direct comparison between the CFD and DCS measurement.

Overall, the DCS and CFD results match closely across the various parameters. 73%, 99%, 100%, and 90% of CFD values fall within the DCS measurement uncertainties for temperature, pressure, $\chi_{H2O}$, and mass flux respectively. Velocity shows more significant disagreement, with only 43% of CFD values falling within DCS measurement uncertainties (which are relatively small). The disagreement is most significant in the two upstream axial locations where DCS measures sharper gradients in the vertical velocity profile than the CFD predicts. However overall, CFD and DCS velocities only differ by 4% magnitude on average.

## 4. Experiment 2: 2D Velocity Profiles in a Combustor

We make the first DCS velocity measurements in a high-speed combustor. These consist of vertical scans before and after the thermal throat. The thermal throat is the location downstream of the flame holder where the flow returns to the supersonic condition. For these measurements, we adjust the optical filter to 6800 to 7000 cm$^{-1}$. This lower wavenumber bound captures more absorption features with high lower-state energies (E"). These features begin to absorb more strongly at the 1500-2000 K temperatures present in the combustor and offer more sensitivity for measurements at these conditions. DCS light is transmitted through the test section in the same way described in the Sections 2 and 3, but with angles of 15° and -15° to the flow corresponding to a 2.8 cm axial span. The two vertical scans were taken at axial positions 1268 and 1478 mm downstream of the isolator entrance, as shown in Fig. 1 (vertical green lines), with vertical measurement spacings of 4-7 mm.

As sustained combustion is not possible with the quartz windows on the combustor, the combustor is ignited in short bursts of 5 seconds, 12 times per measurement condition. DCS measurements for each burst of a single condition were then stitched together using the facility pressure-based flag that indicates

Table 1

Measurement uncertainties by source for measurement at 240 mm axial and 16 mm height position combustion is occurring. Figure 4a shows a comparison of a time trace of the combustion flag against a ~6 Hz temporally resolved $\chi_{H2O}$ fit from the DCS spectra (blue trace). Since there is a transient period before and after stable combustion condition is achieved, we trim 0.6 s (520 spectra at 626 Hz) after and before the on- and off-combustion flags, respectively. This methodology enables approximately 45 s of total averaging for each measurement. An example 45 s averaged absorbance spectrum is shown in Fig. 2.

| Source/Parameter | Velocity | Pressure | Temperature | $\chi_{H2O}$ | Air Mass Flux |
| --- | --- | --- | --- | --- | --- |
| DCS Instrument Accuracy | $2.5 \times 10^{-6}$% | $2.5 \times 10^{-6}$% | -- | -- | $3.5 \times 10^{-6}$% |
| DCS Instrument Precision | 1.8% | 1.5% | 1.0% | 1.4% | 2.5% |
| Beam Angle | 2.3% | 0.1% | 0.0% | 0.5% | 2.3% |
| Background Subtraction | 0.6% | 0.5% | 0.8% | 0.9% | 1.2% |
| Database Uncertainty | 0.0% | 5.4% | 1.4% | 8.3% | 5.6% |
| Total | 3.0% | 5.6% | 1.9% | 8.5% | 6.6% |

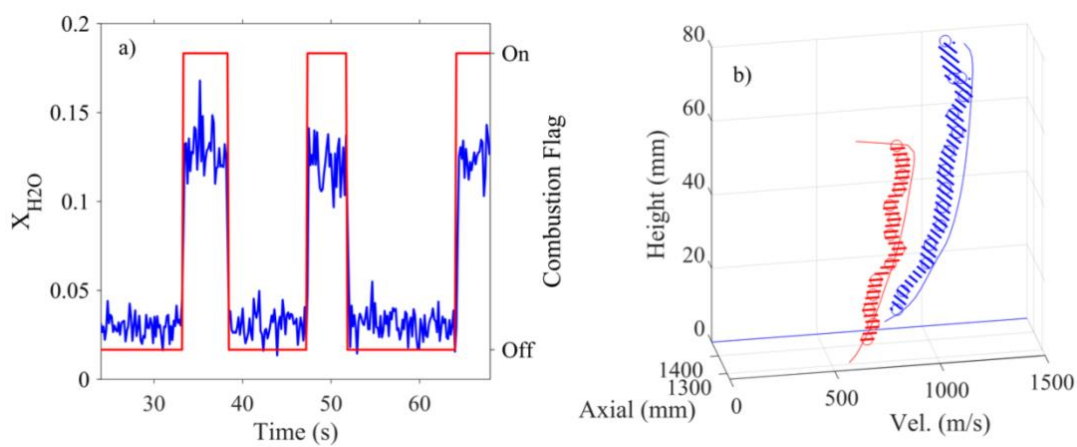

Fig. 4. a) Comparison of ~6 Hz DCS $\chi_{H2O}$ fits (blue) and a time trace of the facility combustion flag (red) for the measurement at 1268 mm axial and 44 mm height in the combustion. b) Comparison of DCS-measurements (markers) and CFD values (lines) of velocity at various heights and axial locations in the isolator. Zero height is defined as the bottom of the isolator and zero axial position is defined as the isolator entrance. Hatched regions indicate the DCS measurement uncertainty bounds.

We use the same 300-1300 K H$_2$O database to fit the combustor data, but with temperature-dependence exponents for pressure-broadening for the highest-temperature H$_2$O features estimated from Antony et al. [36] to get a more accurate lineshape fit for the especially high temperatures (1500-2000 K) experienced in the combustor. While the database fits the measured data well enough to offer low uncertainty velocimetry measurements (which use the differential shift in the position of absorption features between the upstream and downstream beams), the absorption database is not designed for or validated at the 1500-2000 K temperatures present at this combustor location. We therefore do not include measurements of temperature, pressure, $\chi_{H2O}$, and mass flux at this location because we cannot provide a robust estimate of uncertainty for these quantities.

DCS and CFD velocity measurements are shown in Fig. 4b, with the shaded region indicative of the DCS uncertainty. To model the 3D reacting flow in the combustor, the CFD employs a reduced reaction mechanism involving 25 species for vaporized JP-7 fuel developed and validated by [37]. This JP-7 model is based on a detailed mechanism for combustion of large hydrocarbon fuels called HyChem [38,39]. As in the isolator, measurements with low laser light throughput are not included.

While only 24% of CFD velocities fall within the DCS measurement uncertainty, there is a general agreement of ~6% magnitude on average. Both the DCS measurement and CFD predict a relatively flat velocity profile with comparable acceleration through the thermal throat.

An uncertainty analysis performed in the same manner as Section 3 yields uncertainties of 3% to 5% across the combustor DCS velocity measurements. An example of the uncertainty analysis can be seen in Table 2 for the measurement at 1268 mm axial and 48 mm height. There is an additional source of uncertainty in the combustor pertaining to the startup and shutdown combustion transient and the appropriate time to trim from the measurement to remove this transient. This uncertainty is estimated by examining fits from different trim lengths. However, as this uncertainty is not able to be differentiated from the precision (which also contributes to uncertainty when examining fits from different trim lengths), we include both uncertainties under the "precision" in the uncertainty analysis. In contrast to the isolator velocity uncertainty, the database uncertainty for velocimetry is non-negligible in the combustor. Typically, the dual-beam configuration effectively cancels out database error (because it is a differential measurement). However, in the combustor, model database errors at >1500 K are large enough that a fit error arises from the interaction of the noise on the absorbance spectrum and the model error. We estimate this uncertainty by fitting the data with several different databases and find the spread in velocity fits to be 1.3%.

Table 2
Velocity uncertainty by source for combustor measurement at 1268 mm axial position and 48 mm height

| Source/Parameter | Velocity |
|---|---|
| DCS Instrument Accuracy | $2.5 \times 10^{-6}$% |
| DCS Instrument Precision | 3.7% |
| Beam Angle | 1.9% |
| Background Subtraction | 0.6% |
| Database Uncertainty | 1.3% |
| Total | 4.4% |

## 5. Conclusion

We present 2D profiles of temperature, pressure, velocity, $\chi_{H2O}$, and mass flux measured with mode-locked DCS in a ground-test dual-mode ramjet engine. We demonstrate measurements in two different environments in the engine across thermodynamics ranges of 0.6-2 atm pressure, 550-2000 Kelvin temperature, and 500-1200 m/s velocity. The DCS measurements provide 2D profiles which resolve gradients that cannot be detected by wall-mounted or intrusive sensors. CFD calculations of the flow parameters show good agreement with the DCS across oblique shocks in the isolator. Aside from velocity, almost all CFD values fall within the uncertainty of the DCS measurements. DCS measurements of velocity in the combustor demonstrate the effects of the thermal throat. While CFD-derived velocity values across the thermal throat of the engine generally don't fall within the DCS uncertainty, the measurements and CFD are still within 6% on average in magnitude and variation. An analysis considering all major sources of measurement uncertainty demonstrates that the uncertainty for most quantities is driven by non-instrument sources (e.g. absorption models and databases) that can be improved. These results demonstrate the potential of DCS to be a non-intrusive, accurate, and absolute diagnostic for supersonic flows.

## Acknowledgements

This research was sponsored by the Defense Advanced Research Projects Agency (W31P4Q-15-1-0011) the Air Force Office of Scientific Research (FA9550-17-1-0224, FA8650-20-2-2418), and the Air Force Research Laboratory (FA8650-20-2-2418). We thank Steve Enneking, Andrew Baron, Justin Stewart, and the facility operators who worked hard to ensure a smooth experimental campaign in Research Cell 18.